\documentclass[conference, comsoc, 10pt]{IEEEtran}
\IEEEoverridecommandlockouts

\usepackage[
    shortcuts,       
    acronym,         
    nomain,          
    section=chapter, 
    toc=false,        
    style=super,
]{glossaries}
\glsenableentrycount 

\usepackage{siunitx} 
\usepackage{amssymb}
\usepackage{amsmath}
\usepackage{amsfonts}
\usepackage{mathtools}
\usepackage{graphicx}
\usepackage{color}
\usepackage{caption}
\usepackage{subcaption}
\usepackage{cite}
\usepackage{changepage}
\usepackage{adjustbox}
\usepackage{cancel}
\usepackage{tabularx}
\usepackage{makecell}
\usepackage[linesnumbered, ruled, lined]{algorithm2e}
\DontPrintSemicolon
\usepackage{bm}
\usepackage{todonotes}
\usepackage{standalone}
\usepackage{hyperref}
\usepackage[capitalize]{cleveref} 
\usepackage{tikz}
\usepackage{pgfplots}
\pgfplotsset{compat=newest}
\usepackage{aligned-overset}

\usetikzlibrary{shapes.geometric}
\usetikzlibrary{arrows.meta}
\usetikzlibrary{calc}
\usetikzlibrary{spy}

\graphicspath{{images/}} 

\newcommand{\fig}[1]{Fig.~\ref{#1}}    
\newcommand{\secref}[1]{Section~\ref{#1}}

\newcommand{\algoref}[1]{Algorithm~\ref{#1}}

\newcommand{\B}[1]{{\bm{#1}}}

\renewcommand{\d}{\operatorname{d} \!}
\renewcommand{\H}{^{\mathrm{H}}}

\newcommand{\T}{^{\mathrm{T}}}

\newcommand{\eye}{\mathbf{I}}

\newcommand{\C}{\mathbb{C}}

\newcommand{\NC}{\mathcal{N}_{\C}}

\newcommand{\khat}{{\hat{k}}}
\newcommand{\ktilde}{{\tilde{k}}}

\newcommand{\Btheta}{{\B \theta}}
\newcommand{\Bh}{\B h}

\newcommand{\By}{\B y}
\newcommand{\BH}{\B H}

\newcommand{\BY}{\B Y}
\newcommand{\Bn}{\B n}
\newcommand{\BN}{\B N}

\newcommand{\Nrx}{{N_{\mathrm{rx}}}}
\newcommand{\Ntx}{{N_{\mathrm{tx}}}}
\newcommand{\Np}{{N_{\mathrm{p}}}}

\DeclareMathOperator*{\argmin}{\arg\!\min}

\DeclareMathOperator{\E}{\mathbb{E}}

\DeclareMathOperator{\vect}{vec}

\SetCommentSty{footnotesize}

\newcommand{\lineWidth}{1.0pt}
\newcommand{\markSize}{2.2pt}

\pgfplotsset{every axis/.append style={
                    label style={font=\scriptsize},
                    tick label style={font=\scriptsize} 
                    }}

\tikzset{standardoptions/.style={
	line width=\lineWidth,
	mark size=\markSize
}}

\newacronym[nonumberlist]{3gpp}{3GPP}{3rd Generation Partnership Project}
\newacronym[nonumberlist, longplural={angles of arrival}]{aoa}{AOA}{angle of arrival}
\newacronym[nonumberlist]{aod}{AOD}{angle of departure}
\newacronym[nonumberlist]{ae}{AE}{autoencoder}
\newacronym[nonumberlist]{ai}{AI}{artificial intelligence}
\newacronym[nonumberlist]{bn}{BN}{batch normalization}
\newacronym[nonumberlist]{bs}{BS}{base station}
\newacronym[nonumberlist]{awgn}{AWGN}{additive white Gaussian noise}
\newacronym[nonumberlist]{cdl}{CDL}{clustered delay line}
\newacronym[nonumberlist]{ce}{CE}{channel estimation}
\newacronym[nonumberlist]{cme}{CME}{conditional mean estimator}
\newacronym[nonumberlist]{cnf}{CNF}{continuous normalizing flow}
\newacronym[nonumberlist]{cnn}{CNN}{convolutional neural network}
\newacronym[nonumberlist]{cs}{CS}{compressive sensing}
\newacronym[nonumberlist]{csi}{CSI}{channel state information}
\newacronym[nonumberlist]{ddim}{DDIM}{denoising diffusion implicit model}
\newacronym[nonumberlist]{ddpm}{DDPM}{denoising diffusion probabilistic model}
\newacronym[nonumberlist]{dft}{DFT}{discrete Fourier transformation}
\newacronym[nonumberlist]{dl}{DL}{downlink}
\newacronym[nonumberlist]{dm}{DM}{diffusion model}
\newacronym[nonumberlist]{dmce}{DMCE}{diffusion model-based channel estimator}
\newacronym[nonumberlist]{dsm}{DSM}{denoising score matching}
\newacronym[nonumberlist]{elbo}{ELBO}{evidence lower bound}
\newacronym[nonumberlist]{em}{EM}{expectation-maximization}
\newacronym[nonumberlist]{fft}{FFT}{fast Fourier transform}
\newacronym[nonumberlist]{fid}{FID}{Fréchet inception distance}
\newacronym[nonumberlist]{flop}{FLOP}{floating point operation}
\newacronym[nonumberlist]{gan}{GAN}{generative adversarial network}
\newacronym[nonumberlist]{genai}{GenAI}{generative artificial intelligence}
\newacronym[nonumberlist]{gmm}{GMM}{Gaussian mixture model}
\newacronym[nonumberlist]{hvae}{HVAE}{hierarchical variational autoencoder}
\newacronym[nonumberlist]{ifft}{IFFT}{inverse FFT}
\newacronym[nonumberlist]{iid}{i.i.d.}{independently and identically distributed}
\newacronym[nonumberlist]{kl}{KL}{Kullback-Leibler}
\newacronym[nonumberlist]{kpi}{KPI}{key performance indicator}
\newacronym[nonumberlist]{ldm}{LDM}{latent diffusion model}
\newacronym[nonumberlist]{lmmse}{LMMSE}{linear minimum mean squared error}
\newacronym[nonumberlist]{los}{LOS}{line of sight}
\newacronym[nonumberlist]{ls}{LS}{least squares}
\newacronym[nonumberlist]{mimo}{MIMO}{multiple-input multiple-output}
\newacronym[nonumberlist]{miso}{MISO}{multiple-input single-output}
\newacronym[nonumberlist]{ml}{ML}{machine learning}
\newacronym[nonumberlist]{mmd}{MMD}{maximum mean discrepancy}
\newacronym[nonumberlist]{mmse}{MMSE}{minimum mean square error}
\newacronym[nonumberlist]{mmwave}{mmWave}{millimeter wave}
\newacronym[nonumberlist]{mse}{MSE}{mean squared error}
\newacronym[nonumberlist]{mt}{MT}{mobile terminal}
\newacronym[nonumberlist]{nf}{NF}{normalizing flow}
\newacronym[nonumberlist]{nfe}{NFE}{neural function evaluation}
\newacronym[nonumberlist]{nlos}{NLOS}{non-line of sight}
\newacronym[nonumberlist]{nlp}{NLP}{natural language processing}
\newacronym[nonumberlist]{nmse}{NMSE}{normalized mean squared error}
\newacronym[nonumberlist]{nn}{NN}{neural network}
\newacronym[nonumberlist]{ode}{ODE}{ordinary differential equation}
\newacronym[nonumberlist]{ofdm}{OFDM}{orthogonal frequency division multiplex}
\newacronym[nonumberlist]{ofdma}{OFDMA}{orthogonal frequency division multiple access}
\newacronym[nonumberlist]{pdf}{PDF}{probability density function}
\newacronym[nonumberlist]{pmf}{PMF}{probability mass function}
\newacronym[nonumberlist]{quadriga}{QuaDRiGa}{Quasi Deterministic Radio Channel Generator}
\newacronym[nonumberlist]{rbf}{RBF}{radial basis function}
\newacronym[nonumberlist]{resnet}{ResNet}{residual neural network}
\newacronym[nonumberlist]{sbm}{SBM}{score-based generative model}
\newacronym[nonumberlist]{sde}{SDE}{stochastic differential equation}
\newacronym[nonumberlist]{sgd}{SGD}{stochastic gradient descent}
\newacronym[nonumberlist]{silu}{SiLU}{sigmoid linear unit}
\newacronym[nonumberlist]{simo}{SIMO}{single-input multiple-output}
\newacronym[nonumberlist]{snr}{SNR}{signal-to-noise ratio}
\newacronym[nonumberlist]{tdl}{TDL}{tapped delay line}
\newacronym[nonumberlist]{ul}{UL}{uplink}
\newacronym[nonumberlist]{ula}{ULA}{uniform linear array}
\newacronym[nonumberlist]{uma}{UMa}{urban macro-cell}
\newacronym[nonumberlist]{umi}{UMi}{urban micro-cell}
\newacronym[nonumberlist]{ura}{URA}{uniform rectangular array}
\newacronym[nonumberlist]{vae}{VAE}{variational autoencoder}
\newacronym[nonumberlist]{ve}{VE}{variance-exploding}
\newacronym[nonumberlist]{vp}{VP}{variance-preserving}
\newacronym[nonumberlist]{wgan}{WGAN}{Wasserstein generative adversarial network}

\usepackage{tumcolor}

\begin{document}

\title{Enhancements in Score-based Channel Estimation for Real-Time Wireless Systems
\thanks{This work was funded by Huawei Technologies Sweden AB, Stockholm.}
}

\author{Florian Strasser, Marion Bäro, and Wolfgang Utschick\\
\textit{TUM School of Computation, Information and Technology, Technical University of Munich, Germany}\\
E-Mail: \texttt{\{f.strasser, marion.baero, utschick\}@tum.de}}
\maketitle

\begin{abstract}
We propose enhancements to score-based generative modeling techniques for low-latency pilot-based channel estimation in a point-to-point single-carrier multiple-input multiple-output (MIMO) wireless system. Building on recent advances in score-based models, we investigate a specific noise schedule design and sampling acceleration by step-skipping to reduce the number of denoising steps during inference. We additionally propose a single-step signal-to-noise ratio informed denoiser as an extreme case of the step-skipping approach. Our methods achieve significant latency reductions without performance degradation, as demonstrated on a synthetic channel dataset representing an urban macrocell MIMO communications scenario.
\end{abstract}


\section{Introduction}
Accurate \cgls{csi} acquisition is a key factor when scaling up wireless communication systems \cite{Rusek_2013}. However, conventional estimators based on maximum likelihood, \cgls{mmse}, and Bayesian techniques are reaching their limits. Therefore, \cgls{ml} and in particular \cgls{genai} based methods have been proposed due to their great potential in learning the underlying prior distribution of a wireless channel \cite{Huynh_2024}. This distribution covers the site-specific channel characteristics of a \cgls{bs}'s surrounding. In accordingly designed algorithms, exploiting this prior knowledge can significantly improve the channel estimation quality. Hence, various \cgls{genai}-based channel estimation algorithms using, e.g, \cglspl{gmm} \cite{Koller_2022}, \cglspl{gan} \cite{Balevi_2021}, and \cglspl{vae} \cite{Baur_2024} have recently been proposed.

The latest advancements derive channel estimators based on \cglspl{dm} \cite{Ho_2020} and the closely related \cglspl{sbm} \cite{Song_2021_sde}, showcasing great potential. Especially, their prevailing generative abilities compared to, e.g., \cglspl{gan}, together with a stable training process and design flexibility, make them interesting candidates for wireless algorithm development. The major drawback of \cglspl{dm} and \cglspl{sbm} is their high latency during inference due to the iterative sampling procedure. To deploy algorithms that utilize \cglspl{dm}/\cglspl{sbm} in real-time wireless communications systems, we thus have to overcome this limitation.

\subsection{Related Work}
The authors of \cite{Arvinte_2023} introduced a \cgls{sbm}-based framework, incorporating the pilot information in a guidance term during posterior sampling. However, their implementation exhibits high computational complexity. The channel estimator proposed in \cite{Fesl_2024} features a deterministic adaptation of the \cgls{ddpm} \cite{Ho_2020} and a \cgls{snr}-dependent denoising procedure. Also, by exploiting structural properties of the \cgls{mimo} system model, the authors can utilize a lightweight \cgls{nn} implementation, reducing computational complexity. Building on this work, \cite{Zhou_2025} extends the channel estimator to a broader set of problems, specifically under-determined and quantized channel estimation. Additionally, they propose a framework solely using noisy pilots for training, altogether avoiding the necessity of ground-truth channels. In \cite{Ma_2024}, the \cgls{sbm}- and \cgls{ddpm}-based methods are compared to a channel estimator based on a \cgls{ddim} \cite{song_2021_ddim}, which exhibits better performance and also reduces the number of denoising steps. Another recently published work \cite{Kumar_2025}, incorporating many ideas of \cite{Fesl_2024}, improves the estimator's efficiency even more by using the DPM-Solver framework \cite{Lu_2022}. The authors of \cite{Xu_2025} propose to use a Brownian-Bridge \cgls{dm} in an integrated sensing and communications system to estimate the sensing channel.

Apart from channel estimation, \cglspl{dm} and \cglspl{sbm} have also been proposed for semantic communications \cite{Duan_2024, Wu_2024, Zeng_2024}, \cgls{csi} data augmentation \cite{Xu_2023, Lee_2024}, joint channel estimation and data detection \cite{Zilberstein_2024} and end-to-end learning \cite{Kim_2024}. 

\subsection{Contributions}
The main focus of this work is to further reduce the complexity of a \cgls{sbm}-based \cgls{mimo} channel estimation algorithm. We examine improving the iterative denoising procedure of the \cgls{sbm} rather than any improvements in the \cgls{nn} architecture itself. Our contributions can be summarized as follows:
\begin{itemize}
	\item We adapt the channel estimation algorithm proposed in \cite{Fesl_2024} using formulations for a \cgls{sbm} with a \cgls{ve} Gaussian kernel.
	\item We propose two different techniques to accelerate the denoising procedure, i.e., a specific noise schedule design and a step-skipping approach.
	\item We derive a single-step \cgls{snr}-informed channel estimator, which is the extreme case of the step-skipping idea.
	\item We validate the effectiveness of our proposed \cgls{sbm} enhancements with numerical simulations conducted on synthetic channel data.
\end{itemize}

The rest of this work is structured as follows. We introduce the channel estimation problem and the system model in \secref{sec:problem_formulation}. \secref{sec:preliminaries} shortly discusses the \cgls{sbm} framework and explains its usage in the channel estimation algorithm. Enhancements of the \cgls{sbm}-based channel estimator are proposed in \secref{sec:enhancements} and supported with simulation results presented in \secref{sec:simulations}. \secref{sec:conclusion} concludes this work.

\subsection{Notation}
Bold-faced lower-case and upper-case letters denote vectors and matrices, respectively. The matrix $\B F_N$ describes an $N$-dimensional \cgls{dft} matrix and $\eye_M$ is the $M$-dimensional identity matrix. The vectorization (stacking of columns) of a matrix $\B A \in \C^{M \times N}$ is denoted as $\vect({\B A}) \in \C^{MN}$, and $\B A \otimes \B B \in \C^{MP \times NQ}$ is the Kronecker product of $\B A$ and $\B B \in \C^{P \times Q}$. A complex-valued multivariate Gaussian distribution with mean $\B \mu$ and covariance matrix $\B C$ is described by $\NC(\B \mu, \B C)$. The gradient of a function $f$ w.r.t. a variable $\B x$ is written as $\nabla_{\B x} f$.

\section{Problem Formulation}\label{sec:problem_formulation}
We consider a point-to-point single-carrier \cgls{mimo} system with an $\Ntx$-antenna transmitter and an $\Nrx$-antenna receiver. The pilot-based transmission scheme is described by
\begin{equation}\label{eq:system_model_matrix}
	\BY = \BH \B P + \BN,
\end{equation}
where $\BY \in \C^{\Nrx \times \Np}$ comprises the received signals, $\BH \in \C^{\Nrx \times \Ntx}$ describes the unknown channel matrix, $\B P \in \C^{\Ntx \times \Np}$ is the unitary pilot matrix satisfying $\B P \B P\H = \eye_\Ntx$ and $\BN \in \C^{\Nrx \times \Np}$ describes Gaussian noise with columns $\Bn_i \sim \NC (\B 0, \eta^2 \eye)$. The objective is to recover the unknown channel matrix $\BH$ from the observed signal $\BY$. We assume $\Np = \Ntx$, resulting in a fully-determined channel estimation scenario. Rewriting \eqref{eq:system_model_matrix} as 
\begin{equation}\label{eq:system_model_vector}
	\By = (\B P\T \otimes \eye_{\Nrx}) \Bh + \Bn = \B A \Bh + \Bn,
\end{equation}
with $\By = \vect(\BY)$, $\Bh = \vect(\BH)$, $\Bn = \vect(\BN)$ and $\B A=(\B P\T \otimes \eye_{\Nrx})$  enables us to work with vectors instead of matrices. Note that $\B A$ is also unitary since it is derived by the Kronecker product of two unitary matrices.

The channels are assumed to be distributed according to an unknown prior $p(\Bh)$. Assuming $\E_{p}[\Bh] = \B 0$ and $\E_{p}[\| \Bh\|^2] = \Ntx \Nrx$ leads to the \cgls{snr} definition $\mathrm{SNR} (\By) = 1/\eta^2$.

\section{Preliminaries}\label{sec:preliminaries}
\subsection{Score-based Generative Modeling} \label{subsec:sbm}
The \cgls{sbm} considered in this work is based on the \cgls{ve} \cgls{sde} framework introduced in \cite{Song_2021_sde}, which is briefly reviewed in this section. We use the \textit{forward \cgls{sde}}
\begin{equation}\label{eq:forward_sde}
	\d \Bh(t) = \sqrt{\frac{\d [\sigma^2(t)]}{\d t}} \d \B w (t)
\end{equation}
for $t \in [0, T]$ with $\Bh(0) \coloneqq \Bh \sim p(\Bh)$ and a standard Wiener process $\B w(t)$ to continuously diffuse a ground-truth channel sample into Gaussian noise. The noise schedule  $\sigma(t)$ controls the diffusion process and must satisfy the condition $\lim_{t\to0} \sigma(t) =0$. Alternatively, a noisy sample can directly be produced by sampling from the \cgls{ve} Gaussian kernel $q_\sigma(\Bh(t) | \Bh(0)) = \NC(\Bh(t); \Bh(0), \sigma^2(t) \eye)$, such that 
\begin{equation}\label{eq:forward_kernel}
\Bh (t) = \Bh (0) + \sigma(t) \B \epsilon(t)
\end{equation}
using the reparameterization trick with $\B \epsilon(t) \sim \NC(\B 0, \eye)$. This leads to the \cgls{sbm}'s \cgls{snr} definition $\mathrm{SNR}^{\mathrm{SBM}}(t) = 1 / \sigma^2(t)$.

\begin{algorithm}[t]
    \caption{NN training with DSM}
    \label{alg:training}
    \renewcommand{\thealgocf}{}
    \let\oldnl\nl
    \newcommand{\nonl}{\renewcommand{\nl}{\let\nl\oldnl}} 
    \KwIn{Training dataset in beamspace domain $\mathcal{H}_\textrm{train}^\textrm{beam}$, noise variances $\{\sigma_k\}_{k=1}^K$}
        \For{\textnormal{epoch} $\leftarrow 1, \ldots, n_\textnormal{epochs}$}{
            \For{\textnormal{batch} $\leftarrow 1, \ldots, n_\textnormal{batches}$}{
                \For{$i\leftarrow 1, \ldots, \textnormal{batch size}$}{
                    $\Bh_0 \sim \mathcal{H}_\textrm{train}^\textrm{beam}$ \tcp*{draw sample}
                    $k \sim \mathcal{U}\{1, K\}$ \tcp*{draw noise level}
                    $\B z_0 \sim \NC(\B 0, \eye)$ \tcp*{draw noise realization}
                    $\B s^{(i)} \leftarrow - \frac{\B z_0}{\sigma_k}$ \tcp*{calcualate surrogate score}
                    $\Bh_k \leftarrow \Bh_0 + \sigma_k \B z_0$ \tcp*{apply VE Gaussian kernel}
                    $\hat{\B s}_\Btheta^{(i)} \leftarrow \B s_\Btheta(\Bh_k, k)$ \tcp*{estimate score with NN}
                }
                \textnormal{Loss} $\leftarrow \underset{i}{\textnormal{MSE}}(\B s^{(i)}, \hat{\B s}_\Btheta^{(i)})$ \tcp*{compute loss}
                $\Btheta \leftarrow \textnormal{AdamW}(\Btheta, \textnormal{Loss})$ \tcp*{optimization step} 
            }
	}
	\renewcommand{\thealgocf}{\arabic{algocf}} 
\end{algorithm}

The \textit{reverse \cgls{sde}} of \eqref{eq:forward_sde} moving from $t=T$ down to $t=0$ is derived by
\begin{equation}\label{eq:reverse_sde}
	\d \Bh(t) = - \frac{\d [\sigma^2(t)]}{\d t} \nabla_{\Bh (t)}\!\log p(\Bh(t)) +\sqrt{\frac{\d [\sigma^2(t)]}{\d t}} \d \bar{\B w} (t)
\end{equation}
introducing the score function $\nabla_{\Bh (t)}\!\log p(\Bh(t))$ and the reverse-time Wiener process $\bar{\B w} (t)$. Discretizing \eqref{eq:reverse_sde} according to the Euler-Maruyama method \cite{Kloeden_1992} yields
\begin{equation*}\label{eq:reverse_sde_discrete_2}
	\Bh_{k-1} = \Bh_k + (\sigma_k^2 - \sigma_{k-1}^2) \nabla_{\Bh_k}\!\log p(\Bh_k) + \sqrt{\sigma_k^2 - \sigma_{k-1}^2}\B z_k.
\end{equation*}
From the discretization follows that $\Bh(t) \rightarrow \{\Bh_k\}_{k=1}^K$ and $\sigma(t) \rightarrow \{\sigma_k\}_{k=1}^K$, where $K$ is the total number of discretization points. However, the score function in \eqref{eq:reverse_sde_discrete_2} is generally unknown and is thus approximated with a \cgls{nn} $\B s_\Btheta$ parametrized by $\Btheta$ at each discretization step using \cgls{dsm} \cite{Pascal_2011}. The corresponding training procedure is outlined in \algoref{alg:training}. Upon successful training, the \cgls{nn} replaces the score function in \eqref{eq:reverse_sde_discrete_2} to get
\begin{equation}\label{eq:reverse_sde_discrete}
	\Bh_{k-1} = \Bh_k + (\sigma_k^2 - \sigma_{k-1}^2) \B s_\Btheta(\Bh_k, k) + \sqrt{\sigma_k^2 - \sigma_{k-1}^2}\B z_k.
\end{equation}
The resulting iterative sampling procedure can also be interpreted as a Markov chain with Gaussian transition distributions $p_\Btheta(\Bh_{k-1} | \Bh_{k}) = \NC(\Bh_{k-1}; \B \mu_{\Btheta} (\Bh_k, k), \B C_k)$ with means $\B \mu_{\Btheta} (\Bh_k, k) = \Bh_k + (\sigma_k^2 - \sigma_{k-1}^2) \B s_\Btheta(\Bh_k, k)$ and covariance matrices $\B C_k = (\sigma_k^2 - \sigma_{k-1}^2) \eye$.

\subsection{Channel Estimation Algorithm}\label{subsec:channel_estimator}
\begin{algorithm}[t]
	\caption{\cgls{sbm}-based channel estimation}
	\label{alg:sbm_estimator}
	\renewcommand{\thealgocf}{}
	\let\oldnl\nl
	\newcommand{\nonl}{\renewcommand{\nl}{\let\nl\oldnl}}
	\KwIn{Pre-trained SBM $s_\Btheta$, $\{\sigma_k\}_{k=1}^K$, $\By$, $\eta^2$, $\B A$ }
	$\hat{\B h} \leftarrow  \B A\H \B y$ \tcp*{decorrelate pilot matrix}
	$\hat{\Bh} \leftarrow (\B F_\Ntx \otimes \B F_\Nrx) \hat{\Bh}$ \tcp*{beamspace transformation}
	$\khat= \argmin_k |\eta^2 - \sigma_k^2|$ \tcp*{identify initial step}
	$\hat{\Bh}_{\khat} \leftarrow \hat{\B h}$ \tcp*{initialize \acrshort{sbm} sampling process}
	\For{$k=\khat$ \textnormal{\textbf{down to}} $1$}{
		$\hat{\B h}_{k-1} \leftarrow \hat{\Bh}_k + (\sigma_k^2 - \sigma_{k-1}^2) \B s_\Btheta(\Bh_k, k) $\;
	}
	$\hat{\B h} \leftarrow  (\B F_\Ntx\H \otimes \B F_\Nrx\H) \hat{\B h}_{0}$ \tcp*{inverse beamspace transformation}
	\renewcommand{\thealgocf}{\arabic{algocf}} 
\end{algorithm}

We build on the \cgls{dm}-based channel estimation algorithm proposed in \cite{Fesl_2024}, which combines an \cgls{snr}-informed estimator with a lightweight \cgls{cnn}. We also train our \cgls{sbm} in the beamspace domain by applying a Kronecker \cgls{dft} to the spatial channels, thereby exploiting structural properties of wireless signals. Also, we adopt their \cgls{cnn} implementation and the \cgls{snr}-dependent estimation procedure. However, the authors of \cite{Fesl_2024 } employ a \cgls{vp} \cgls{dm} architecture instead of \cgls{ve} Gaussian kernels. Therefore, some derivations have to be slightly adapted for this work. 

\algoref{alg:sbm_estimator} concisely presents the adapted \cgls{sbm}-based channel estimation algorithm. After the initial decorrelation step of the pilot observation $\By$, we directly conduct the beamspace transformation without a preceding normalization step. This can be omitted due to the equivalence between the decorrelated pilot observation and the \cgls{sbm}'s \cgls{ve} Gaussian kernel. The appropriate initial \cgls{sbm} step is found by identifying the \cgls{sbm} variance $\sigma_k^2$ closest to the received observation's variance $\eta^2$ in line 3. The algorithm then enters the denoising procedure, where it iteratively applies \eqref{eq:reverse_sde_discrete} but without the stochastic term $\sqrt{\sigma_k^2 - \sigma_{k-1}^2}\B z_k$ in line 6. Including this term is reasonable for generative tasks but not for point-wise estimation \cite{Fesl_2024_2}. Upon completion, the channel estimate in the beamspace domain is returned and transformed back into the spatial domain to finish the algorithm.

\section{Enhancements in Score-based Modeling}\label{sec:enhancements}
The main bottleneck in \algoref{alg:sbm_estimator} is the number of required denoising steps $\khat$ as each step $k$ requires a costly \cgls{nfe} of $\B s_\Btheta(\Bh_k, k)$. A naive approach to reducing $\khat$ is to reduce the total number of discretization points $K$. However, a drawback of this idea is that the worsened resolution of the discretized \cgls{sde} is affecting the estimator's performance \cite[Figure 4]{Fesl_2024}. The following two techniques introduce new approaches to reducing denoising steps.
\subsection{Noise Schedule Design}\label{subsec:noise_schedule}
We consider the commonly used exponential noise schedule $\sigma(t)$, which is discretized by the power-exponentiated geometric sequence
\begin{equation}\label{eq:noise_schedule}
	\sigma_k = \sigma_\mathrm{min}\left(\frac{\sigma_\mathrm{max}}{\sigma_\mathrm{min}}\right)^{(\frac{k-1}{K-1})^\gamma}
\end{equation}
with the parameter $\gamma$ controlling the schedule's steepness. The minimum/maximum variances are derived from the maximum/minimum \cgls{snr} the \cgls{sbm} should cover via the relation
\begin{equation}\label{eq:sbm_snr_relation}
	\sigma_{\text{min/max}} = \sqrt{\frac{1}{\mathrm{SNR}^{\mathrm{SBM}}_{\text{max/min}}}}.
\end{equation}

From \algoref{alg:sbm_estimator}, line 3, we can infer that the noise schedule directly influences $\khat$ and thus affects the algorithm's time complexity. For $\gamma < 1$, the sequence in \eqref{eq:noise_schedule} grows faster and therefore fewer denoising steps $\khat$ compared to a standard geometric sequence with $\gamma = 1$ are required for the same $\eta^2$. However, one has to take care that the \cgls{sbm}'s \cgls{snr} resolution does not get too coarse for high \cglspl{snr}.

\subsection{Accelerated Denoising}\label{subsec:accelerated}

\begin{algorithm}[t]
	\caption{Single-Step Channel Estimation}
	\label{alg:sbm_accelerated}
	\renewcommand{\thealgocf}{}
	\let\oldnl\nl
	\newcommand{\nonl}{\renewcommand{\nl}{\let\nl\oldnl}}
	\KwIn{Pre-trained SBM $s_\Btheta$, $\{\sigma_k\}_{k=1}^K$, $\By$, $\eta^2$, $\B A$ }
	$\hat{\B h} \leftarrow  \B A\H \B y$ \tcp*{decorrelate pilot matrix}
	$\hat{\Bh} \leftarrow (\B F_\Ntx \otimes \B F_\Nrx) \hat{\Bh}$ \tcp*{beamspace transformation}
	$\khat= \argmin_k |\eta^2 - \sigma_k^2|$ \tcp*{identify \textcolor{blue}{noise level}}
	$\hat{\Bh}_{\khat} \leftarrow \hat{\B h}$ \tcp*{initialize \textcolor{blue}{denoising}}
	$\hat{\B h}_{\textcolor{blue}{0}} \leftarrow \hat{\Bh}_{\textcolor{blue}{\khat}} + \sigma_{\textcolor{blue}{\khat}}^2 \B s_\Btheta(\hat{\Bh}_{\textcolor{blue}{\khat}}, \textcolor{blue}{\khat}) $ \textcolor{blue}{\tcp*{single-step denoising}}
	$\hat{\B h} \leftarrow  (\B F_\Ntx\H \otimes \B F_\Nrx\H) \hat{\B h}_0$ \tcp*{inverse beamspace transformation}
	\renewcommand{\thealgocf}{\arabic{algocf}} 
\end{algorithm}

The method to speed up the algorithm even more introduced in this section is based on the general idea of skipping certain latent variables of a \cgls{dm} during sampling as proposed in, e.g., \cite{song_2021_ddim, Lu_2022}. In our \cgls{sbm}-based framework, the adapted deterministic denoising procedure incorporating this approach is inferred from \eqref{eq:reverse_sde_discrete} and given by
\begin{equation}\label{eq:accelerated_reverse}
	\hat{\Bh}_{k-\Delta} = \Bh_k + (\sigma_k^2 - \sigma_{k-\Delta}^2) \B s_\Btheta(\Bh_k, k),
\end{equation}
where we introduced the step variable $\Delta \in \mathbb{N}$. The case $\Delta=1$ exactly resembles the standard method discussed in \secref{subsec:channel_estimator}. For $\Delta>1$, we skip certain points of the original discretization and, as a consequence, enlarge the step-size given by $\sigma_k^2 - \sigma_{k-\Delta}^2$ since $\{\sigma_k\}_{k=1}^K$ is a monotonically increasing sequence. This effect can also be interpreted as a coarser discretization of the \textit{reverse \cgls{sde}} during the denoising procedure. It is important to mention that we still maintain the fine resolution of the \cgls{sde} for identifying the initial step $\khat$.

In \algoref{alg:sbm_estimator}, line 6 has to be replaced with \eqref{eq:accelerated_reverse} to realize the accelerated denoising. Also, we have to ensure that the loop is indexed by $k = {\khat, \khat-\Delta, \khat-2\Delta, \ldots, \khat \mod \Delta}$. In the last denoising step, we set $\sigma_{k-\Delta}=0$ to completely remove the remaining noise in the final iteration. The number of required denoising steps and thereby the algorithm's complexity decreases to $\ktilde_{\Delta} = \lceil \frac{\khat}{\Delta}\rceil$.
\vspace{0.3cm}

\noindent\textbf{Single-Step SNR-informed Denoiser.} In the extreme case of $\Delta = \Delta_{\mathrm{max}} = K$, the number of denoising steps reduces to a single step
\begin{equation}\label{eq:single_step}
	\hat{\Bh}_0 = \Bh_k + \sigma_k^2 \B s_\Btheta (\Bh_k, k)
\end{equation}
which completely removes the denoising loop as depicted in \algoref{alg:sbm_accelerated}. The changes compared to \algoref{alg:sbm_estimator} are highlighted in \textcolor{blue}{blue}. This algorithm rather resembles a \cgls{snr}-informed \cgls{nn}-based denoiser than a framework using a \cgls{sbm}.

To better understand how the single-step denoiser works, recall that in \cgls{dsm}, the \cgls{nn} is trained to estimate the score function of the diffused channel data conditioned on a ground-truth sample $\Bh_0$. This can also be reformulated as
\begin{equation}
	\B s_\Btheta(\Bh_k, k) \approx \nabla_{\Bh_k} q_{\sigma}(\Bh_k | \Bh_0) = - \frac{1}{\sigma_k} \B \epsilon_k,
\end{equation}
where $\B \epsilon_k$ is the noise realization used to generate $\Bh_k$. Therefore, during training, instead of directly estimating the score, we train a \cgls{nn} $ \B \epsilon_\Btheta(\Bh_k, k)$ to approximate $\B \epsilon_k$ and set the learned score to $\B s_\Btheta(\Bh_k, k) = - \frac{1}{\sigma_k} \B \epsilon_\Btheta (\Bh_k, k)$. Incorporating these reformulations into \eqref{eq:single_step} yields
\begin{equation}
	\hat{\Bh}_0 = \Bh_k - \sigma_k \B \epsilon_\Btheta (\Bh_k, k).
\end{equation}
This can be interpreted as the approximated inverse of the \cgls{ve} Gaussian forward kernel expressed in \eqref{eq:forward_kernel}. Thus, adding the scaled estimated score in \eqref{eq:single_step} can be interpreted as a direct denoising step — effectively identifying and removing the added noise in a single step.

\section{Numerical Results and Discussion}\label{sec:simulations}
\subsection{Synthetic Channel Data}
We work with version 2.6 of the \cgls{quadriga} channel simulator \cite{Jaeckel_2014} and consider a \cgls{uma} scenario with a strong \cgls{los} path and additional multi-path components. The \cgls{bs} is placed at a height of $\SI{25}{m}$ and equipped with a \cgls{ula} consisting of $\Nrx = 64$ antennas. It covers a sector of $120^\circ$ in which \cglspl{mt} are placed at random with distances ranging between $\SI{35}{m}$ and $\SI{500}{m}$. Each \cgls{mt} itself consists of a \cgls{ula} equipped with $\Ntx = 16$ antennas. We consider half-wavelength antenna spacing for all \cglspl{ula} and the respective channels are simulated at a frequency of $\SI{6}{GHz}$ for a single snapshot.
Although we choose a large number of antennas at the \cglspl{mt} in this work, the characteristics identified in the simulation results also hold for fewer antennas, and especially also in the case of a \cgls{simo} system with $\Ntx=1$.
A post-processing step removes the effective path gain for further processing according to  \cite{Jaeckel_2014}. Knowledge of the effective path gain is a viable assumption as it can be estimated with small efforts in a real communications system. In total, this generates $1.2 \cdot 10^5$ channel realizations $\BH_m \sim p(\BH)$ for different \cgls{mt} locations in the scenario, which are split up in sets of $M_\text{train} = 10^5$,  $M_\text{val} = 10^4$ and  $M_\text{test} = 10^4$ training, validation and test samples, respectively.

\subsection{Score-based Model Implementation}
The \cgls{sbm} is designed upon the noise schedule in \eqref{eq:noise_schedule} with a total of $K=100$ discretization steps. The maximum and minimum \cgls{snr} covered by the \cgls{sbm} are set to $\mathrm{SNR}^{\mathrm{SBM}}_{\text{max}} = \SI{40}{dB}$ and $\mathrm{SNR}^{\mathrm{SBM}}_{\text{min}} = \SI{-22}{dB}$, resulting in minimum and maximum noise variances of $\sigma_\mathrm{min} = 0.01$ and $\sigma_\mathrm{max} \approx 12.6$ according to \eqref{eq:sbm_snr_relation}. If not stated otherwise, $\gamma=1.0$ and $\Delta=1$ hold.

As already stated in \secref{subsec:channel_estimator}, we reuse the \cgls{cnn} implementation of \cite{Fesl_2024} without performing any specific hyperparameter tuning. The architecture combines a lightweight 2D-\cgls{cnn} with a positional Transformer sinusoidal position embedding of step $k$. This enables parameter sharing of the \cgls{nn} across all discretization steps. The \cgls{cnn} is trained with the \textit{AdamW} optimizer \cite{Loshchilov_2019_AdamW} using a batch size of $128$, an adaptive learning rate, and a stopping criterion depending on the validation loss evolution over training epochs. For each simulation, we train $5$ randomly initialized \cglspl{nn} and pick the one exhibiting the lowest validation loss after training for further use in channel estimation inference. We refer to our publicly available \href{https://github.com/StrasserFlorian/enhanced_sbm_estimator}{GitHub repository} for further implementation details.\footnote{\url{https://github.com/StrasserFlorian/enhanced_sbm_estimator}}

\subsection{Baseline Estimators}
The proposed \cgls{sbm}-based channel estimator is compared to the following baseline algorithms.

The \cgls{ls} estimate denoted as "LS" solves the optimization problem $\hat{\Bh}_\mathrm{LS} = \argmin_{\Bh} \|\By - \B A \Bh \|^2$ and can be derived as $\hat{\Bh}_\mathrm{LS} = \B A\H \By$ as $\B A$ is assumed to be unitary. We also evaluate a \cgls{lmmse} based estimate denoted as "SCov-LMMSE" with a global sample covariance matrix $\B C = \frac{1}{M_\text{train}} \sum_{m=1}^{M_\text{train}} \Bh_m \Bh_m\H$ as $\hat{\Bh}_\textrm{Scov-LMMSE} = \B C (\B C + \eta^2 \eye)^{-1}\By$.

As prior-aided baselines, we choose two different \cgls{gmm}-based estimators for comparison. The first version, denoted as "GMM" and proposed in \cite{Koller_2022}, fits a \cgls{gmm} to the ground-truth \cgls{csi} data such that it approximates the underlying distribution well. The learnt means and covariances of the Gaussian components are then exploited in an adapted \cgls{lmmse} estimator. The second version, denoted as "GMM-kron" and proposed in \cite{Fesl_2022}, follows the same reasoning but trains two separate \cglspl{gmm} on the receiver and transmitter side, respectively. Afterwards, covariance matrices of both \cglspl{gmm} are combined with the Kronecker product to build the final covariance matrices used in the adapted \cgls{lmmse}.

\subsection{Performance Criterion}
We opt for the \cgls{nmse} metric to assess the channel estimator's performance. Therefore, for each noise level, a noisy observation of each channel sample $\Bh_m$ in the test set is created according to \eqref{eq:system_model_matrix}. Then, a channel estimate $\hat{\Bh}_m$ is computed with one of the considered estimators. The \cgls{nmse} for each \cgls{snr} level is evaluated as
\begin{equation}
	\mathrm{NMSE} = \frac{1}{\Nrx \Ntx M_\text{test}} \sum_{m=1}^{M_\text{test}}\|\Bh_m - \hat{\Bh}_m \|_2^2.
\end{equation} 
It is important to note that our \cgls{sbm}-based channel estimator does not require retraining for different \cglspl{snr} but operates on the whole \cgls{snr} range.

\begin{figure}[t]
	\centering 
	
	\begin{minipage}{\columnwidth}
		\centering
		\begin{tikzpicture}
			\begin{axis}
				[width=0.98\columnwidth,
				height=0.6\columnwidth,
				xtick={-15, -10, -5, 0, 5, 10, 15, 20},
				xmin=-15, 
				xmax=20,
				xlabel={SNR [dB]},
				ymin= 0,
				ymax=100,
				ylabel= {$\khat$}, 
				ylabel shift = 0.0cm,
				grid = both,
				legend columns = 2,
				legend style={at={(1.0,1.0)}, anchor=north east, font=\scriptsize},
				legend entries={
					$\gamma=0.2$,
					$\gamma=0.6$,
					$\gamma=1.0$,
					$\gamma=1.6$,
				}
				]
				
				\addplot[standardoptions, color=blue, mark=pentagon, mark repeat=8, mark phase=1]
				table[x=SNRs02, y=step, col sep=semicolon]
				{data/noise_schedule_steps.csv};
				
				\addplot[standardoptions, color=red, mark=o, mark repeat=8, mark phase=1]
				table[x=SNRs06, y=step, col sep=semicolon]
				{data/noise_schedule_steps.csv};
				
				\addplot[standardoptions, color=black, mark=square, mark repeat=8, mark phase=1]
				table[x=SNRs10, y=step, col sep=semicolon]
				{data/noise_schedule_steps.csv};
				
				\addplot[standardoptions, color=TUMOrange, mark=triangle, mark repeat=8, mark phase=1]
				table[x=SNRs16, y=step, col sep=semicolon]
				{data/noise_schedule_steps.csv};

			\end{axis}
		\end{tikzpicture}
		\caption{Required number of denoising steps for varying $\gamma$.}
		\label{fig:steps_noise_schedule}
	\end{minipage}
	
	\vspace{0.5cm}
	
	\begin{minipage}{\columnwidth}
		\centering
		\begin{tikzpicture}[spy using outlines={rectangle, magnification=2.0, width=2.0cm, height=1.3cm, connect spies}]
			\begin{axis}
				[width=0.98\columnwidth,
				height=0.6\columnwidth,
				xtick=data,
				xmin=-15, 
				xmax=20,
				xlabel={SNR [dB]},
				ymode = log, 
				ymin= 4*1e-3,
				ymax=0.5,
				ylabel= {NMSE}, 
				ylabel shift = 0.0cm,
				grid = both,
				legend columns = 2,
				legend style={at={(0.0,0.0)}, anchor=south west, font=\scriptsize},
				legend entries={
					$\gamma=0.2$,
					$\gamma=0.6$,
					$\gamma=1.0$,
					$\gamma=1.6$,
				}
				]
				
				\addplot[standardoptions, color=blue, mark=pentagon]
				table[x=SNR, y=0, col sep=semicolon]
				{data/noise_schedule_nmses.csv};
				
				\addplot[standardoptions, color=red, mark=o]
				table[x=SNR, y=1, col sep=semicolon]
				{data/noise_schedule_nmses.csv};
				
				\addplot[standardoptions, color=black, mark=square]
				table[x=SNR, y=2, col sep=semicolon]
				{data/noise_schedule_nmses.csv};
				
				\addplot[standardoptions, color=TUMOrange, mark=triangle]
				table[x=SNR, y=3, col sep=semicolon]
				{data/noise_schedule_nmses.csv};
				
				\coordinate (spypoint) at (axis cs:5,0.055);    
				\coordinate (magnifyglass) at (axis cs:14,0.16); 
				\spy[thick, rectangle, draw] on (spypoint) in node [fill=white] at (magnifyglass);
				
			\end{axis}
		\end{tikzpicture}
		\caption{\acrshort{nmse} performance for varying $\gamma$.}
		\label{fig:nmse_noise_schedule}
	\end{minipage}
\end{figure}

\begin{figure}[t]
	\centering
	
	\begin{minipage}{\columnwidth}
		\centering
		\begin{tikzpicture}
			\begin{axis}
				[width=0.98\columnwidth,
				height=0.6\columnwidth,
				xtick={-15, -10, -5, 0, 5, 10, 15, 20},
				xmin=-15, 
				xmax=20,
				xlabel={SNR [dB]},
				ymin= 0,
				ymax=100,
				ylabel= {$\ktilde_\Delta$}, 
				ylabel shift = 0.0cm,
				grid = both,
				legend columns = 2,
				legend style={at={(1.0,1.0)}, anchor=north east, font=\scriptsize},
				legend entries={
					$\Delta=1$,
					$\Delta=2$,
					$\Delta=4$,
					$\Delta=8$,
					$\Delta=16$,
					$\Delta=\Delta_\mathrm{max}$
				}
				]
				
				\addplot[standardoptions, color=black, mark=square, mark repeat=8, mark phase=1]
				table[x=SNR, y=Delta01, col sep=semicolon]
				{data/accelerated_steps.csv};
				
				\addplot[standardoptions, color=red, mark=o, mark repeat=8, mark phase=1]
				table[x=SNR, y=Delta02, col sep=semicolon]
				{data/accelerated_steps.csv};
				
				\addplot[standardoptions, color=TUMOrange, mark=triangle, mark repeat=8, mark phase=1]
				table[x=SNR, y=Delta04, col sep=semicolon]
				{data/accelerated_steps.csv};
				
				\addplot[standardoptions, color=teal,mark=diamond, mark repeat=8, mark phase=1]
				table[x=SNR, y=Delta08, col sep=semicolon]
				{data/accelerated_steps.csv};
				
				\addplot[standardoptions, color=violet, mark=star, mark repeat=8, mark phase=1]
				table[x=SNR, y=Delta16, col sep=semicolon]
				{data/accelerated_steps.csv};
				
				\addplot[standardoptions, color=blue, mark=pentagon, mark repeat=8, mark phase=1]
				table[x=SNR, y=DeltaMax, col sep=semicolon]
				{data/accelerated_steps.csv};

			\end{axis}
		\end{tikzpicture}
		\caption{Required number of denoising steps for varying $\Delta$.}
		\label{fig:steps_accelerated}
	\end{minipage}
	
	\vspace{0.5cm}
	
	\begin{minipage}{\columnwidth}
		\centering
		\begin{tikzpicture}[spy using outlines={rectangle, magnification=2.0, width=2.0cm, height=1.3cm, connect spies}]
			\begin{axis}
				[width=0.98\columnwidth,
				height=0.6\columnwidth,
				xtick=data,
				xmin=-15, 
				xmax=20,
				xlabel={SNR [dB]},
				ymode = log, 
				ymin= 4*1e-3,
				ymax=0.5,
				ylabel= {NMSE}, 
				ylabel shift = 0.0cm,
				grid = both,
				legend columns = 2,
				legend style={at={(0.0,0.0)}, anchor=south west, font=\scriptsize},
				legend entries={
					$\Delta=1$,
					$\Delta=2$,
					$\Delta=4$,
					$\Delta=8$,
					$\Delta=16$,
					$\Delta=\Delta_\mathrm{max}$,
				}
				]
				
				\addplot[standardoptions, color=black, mark=square]
				table[x=SNR, y=1, col sep=semicolon]
				{data/accelerated_nmses.csv};
				
				\addplot[standardoptions, color=red, mark=o]
				table[x=SNR, y=2, col sep=semicolon]
				{data/accelerated_nmses.csv};
				
				\addplot[standardoptions, color=TUMOrange, mark=triangle]
				table[x=SNR, y=4, col sep=semicolon]
				{data/accelerated_nmses.csv};
				
				\addplot[standardoptions, color=teal, mark=diamond]
				table[x=SNR, y=8, col sep=semicolon]
				{data/accelerated_nmses.csv};
				
				\addplot[standardoptions, color=violet, mark=star]
				table[x=SNR, y=16, col sep=semicolon]
				{data/accelerated_nmses.csv};
				
				\addplot[standardoptions, color=blue, mark=pentagon]
				table[x=SNR, y=max, col sep=semicolon]
				{data/accelerated_nmses.csv};
				
				\coordinate (spypoint) at (axis cs:5,0.055);    
				\coordinate (magnifyglass) at (axis cs:14,0.16); 
				\spy[thick, rectangle, draw] on (spypoint) in node [fill=white] at (magnifyglass);
				
			\end{axis}
		\end{tikzpicture}
		\caption{\acrshort{nmse} performance for varying $\Delta$.}
		\label{fig:nmse_accelerated}
	\end{minipage}
\end{figure}

\subsection{Noise Schedule Design Evaluation}
The effect of choosing different noise schedules on the number of required denoising steps is depicted in \fig{fig:steps_noise_schedule} for different choices of the parameter $\gamma$. Because the noise schedule is an exponential function and we plot the \cglspl{snr} on a logarithmic scale, a linear decrease of $\khat$ can be seen for the regular geometric sequence with $\gamma=1$ in black. As proposed in \secref{subsec:noise_schedule} and validated in \fig{fig:steps_noise_schedule}, decreasing $\gamma$ significantly decreases $\khat$ for every $\mathrm{SNR}(\By)$. We also include a simulation for $\gamma>1$ to investigate the effect on the performance when increasing the number of denoising steps. 

\fig{fig:nmse_noise_schedule} depicts the \cgls{nmse} for the different choices of $\gamma$. There is almost no degradation in performance when decreasing $\gamma$, only in the high-\cgls{snr} regime. The deterioration stems from the \cgls{sde} discretization being too coarse in the low-noise region for small values of $\gamma$. Also, as the graph for $\gamma=1.6$ is on par with the others, there is no benefit in increasing the number of denoising steps by adapting the noise schedule. This indicates that the algorithm can denoise a pilot observation well as long as we maintain a good enough \cgls{sde} discretization.

\subsection{Accelerated Denoising Evaluation}
Even more impressive is the reduction of denoising steps with the accelerated denoising approach, cf. \secref{subsec:accelerated}, as depicted in \fig{fig:steps_accelerated}. The black curve again resembles the baseline implementation, where every discretization step is conducted in the denoising procedure. With increasing $\Delta$, we can accelerate the denoising procedure by roughly a factor of $\Delta$ until only a single step is required for $\Delta=\Delta_\mathrm{max}$ independent of the \cgls{snr}. 

Surprisingly, increasing the step variable $\Delta$ does not worsen the \cgls{nmse}.
\fig{fig:nmse_accelerated} showcases that it even improves the channel estimation performance by a small but significant margin, not stemming from computational instabilities, over the whole \cgls{snr} range.
Especially, in the mid-\cgls{snr} range, the improvement is clearly visible. This result seems counterintuitive at first sight; however, we make the following hypothesis describing the effect: We refer to \cite[Fig. 4]{Fesl_2024}, where the authors show the evolution of the \cgls{nmse} performance during the denoising procedure for different \cgls{snr} levels. However, one artifact not discussed in the paper is a rise in the \cgls{nmse} at the final steps of the estimation procedure after the \cgls{nmse} has reached its minimum. Although we still do not fully understand this effect, we claim that increasing $\Delta$ mitigates this effect, which explains the counterintuitive \cgls{nmse} improvement in \fig{fig:nmse_accelerated}.

\subsection{Comparison to Baselines}

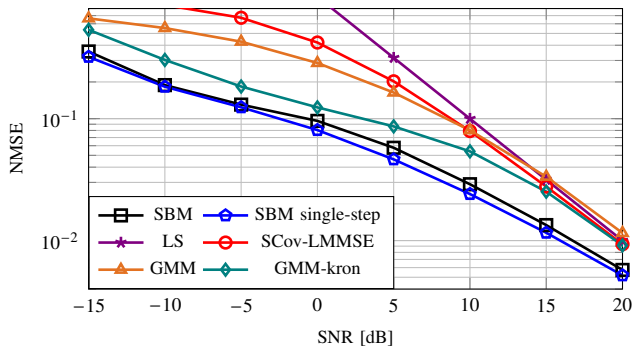
\begin{figure}[t]
		\centering
		\begin{tikzpicture}
			\begin{axis}
				[width=0.98\columnwidth,
				height=0.6\columnwidth,
				xtick=data,
				xmin=-15, 
				xmax=20,
				xlabel={SNR [dB]},
				ymode = log, 
				ymin= 4*1e-3,
				ymax=0.8,
				ylabel= {NMSE}, 
				ylabel shift = 0.0cm,
				grid = both,
				legend columns = 2,
				legend style={at={(0.0,0.0)}, anchor=south west, font=\scriptsize},
				legend entries={
					SBM,
					SBM single-step,
					LS,
					SCov-LMMSE,
                    GMM,
                    GMM-kron
				}
				]
				
				\addplot[standardoptions, color=black, mark=square]
				table[x=SNR, y=1, col sep=semicolon]
				{data/accelerated_nmses.csv};
				
				\addplot[standardoptions, color=blue, mark=pentagon]
				table[x=SNR, y=max, col sep=semicolon]
				{data/accelerated_nmses.csv};

                    \addplot[standardoptions, color=violet, mark=star]
				table[x=SNR, y=LS, col sep=comma]
				{data/sample_cov_ls.csv};

                    \addplot[standardoptions, color=red, mark=o]
				table[x=SNR, y=lmmse_glob, col sep=comma]
				{data/sample_cov_ls.csv};

                    \addplot[standardoptions, color=TUMOrange, mark=triangle]
				table[x=SNR, y=gmm_full, col sep=comma]
				{data/test_results_gmm.csv};

                    \addplot[standardoptions, color=teal, mark=diamond]
				table[x=SNR, y=gmm_kron, col sep=comma]
				{data/test_results_gmm_kron.csv};

			\end{axis}
		\end{tikzpicture}
		\caption{Comparison with baseline estimators.}
		\label{fig:baselines}
\end{figure}
\fig{fig:baselines} depicts the performance of different channel estimators measured in terms of the \cgls{nmse}. The LS and SCov-LMMSE estimators, as two conventional methods, are barely able to reconstruct the channel well, thereby emphasizing the need for more advanced estimation algorithms. The \cgls{gmm}-based approaches and especially the Kronecker version thereof significantly improve the performance in the low-to-mid \cgls{snr}-range while converging to the classical estimators' curves for higher \cglspl{snr}. 

The estimator denoted as "SBM" in \fig{fig:baselines} implements the \cgls{sbm}-based channel estimator from \algoref{alg:sbm_estimator} without step-skipping ($\Delta=1$) and the regular exponential noise schedule ($\gamma=1.0$). It therefore requires the regular number of denoising steps given by $\hat{k}$ and thus is computationally complex. However, it is capable of outperforming the other baselines over the whole \cgls{snr}-range by roughly \SI{5}{dB}. The second \cgls{sbm}-based method, denoted as "SBM single-step", implements the single-step \cgls{snr}-informed denoiser ($\Delta = \Delta_\mathrm{max}$) from \algoref{alg:sbm_accelerated}. As already discussed in the previous section and shown in Figs. \ref{fig:steps_accelerated} and \ref{fig:nmse_accelerated}, this technique is capable of further improving the \cgls{nmse} while drastically reducing the computational complexity. 

\section{Conclusion and Outlook} \label{sec:conclusion}
This paper introduced two methods to accelerate the denoising procedure of a \cgls{sbm}-based \cgls{mimo} channel estimator. By carefully adapting the \textit{forward \cgls{sde}'s} noise schedule, we significantly reduced the number of denoising steps with only minor performance losses. An even faster acceleration was achieved by skipping certain steps in the denoising procedure, which in the extreme case results in the single-step denoiser. This approach even improves the estimation performance due to the mitigation of an artifact in a previously proposed \cgls{dm}-based channel estimator.

Further research should clarify the theoretical optimality guarantees of the proposed denoiser and identify its limitations. Also, investigating the performance for different wireless communications scenarios is essential for the algorithm's generalization capabilities. Regarding this point, an interesting research question not considered in this work is, e.g., what effect a rotation of the \cgls{mt} \cglspl{ula} has on the channel estimation performance. As a final suggestion, the proposed estimator should be adapted to operate in a \cgls{ofdm}ed MIMO system for actual deployment in real-world communications scenarios.

\bibliographystyle{IEEEtran}
\bibliography{IEEEabrv,biblio}

\end{document}